\documentclass[11pt]{article}
\usepackage[english]{babel}
\usepackage[T1]{fontenc}
\usepackage{amsmath,amsthm,amsfonts,amssymb,amsopn,amscd, mathrsfs}
\usepackage{cite}
\usepackage{hyperref}

\newlength{\dinwidth}
\newlength{\dinmargin}
\newcommand{\EE}{E_{\#}}

\newcommand{\yb}{\boldsymbol y}
\newcommand{\PP}{E}
\newcommand{\e}{\mathrm e}
\newcommand{\vv}{\om}
\newcommand{\LLL}{\mathcal L}

\newcommand{\Sp}{\te{Sp}}

\newcommand{\Ri}{+}    
\newcommand{\Le}{-}  

\newcommand{\cc}{\mathrm{c}}

\renewcommand{\t}{s}
\newcommand{\pb}{\boldsymbol p}   
\newcommand{\xb}{\boldsymbol x}   
\newcommand{\Pb}{\boldsymbol P }
\renewcommand{\r}{{\rm{R}}}
\renewcommand{\l}{{\rm{L}}}

\newcommand{\hilk}{\mathcal{K}} 

\newcommand{\U}{V}

\newcommand{\I}{\mathfrak{I}}
\newcommand{\J}{\mathfrak{J}}

\newcommand{\cl}{\te{cl}}
\newcommand{\Span}{\te{Span}}

\newcommand{\tf}{\tilde f}

\newcommand{\De}{\Delta}

\newcommand{\ti}{\tilde}
\newcommand{\slim}{\te{s}\textrm{-}\lim}
\newcommand{\pout}{\overset{\tout}{\times}}

\newcommand{\tout}{\te{ out}}
\newcommand{\Phip}{\Phi^{\tout}_+}
\newcommand{\Phim}{\Phi^{\tout}_-}
\newcommand{\ph}{\phantom}
\newcommand{\x}{x}

\newcommand{\Om}{\Omega}

\newcommand{\h}{\fr{1}{2}}
\newcommand{\te}{\mathrm}

\newcommand{\de}{\delta}

\newcommand{\nat}{\mathbb{N}}
\newcommand{\hil}{\mathcal{H}}
\newcommand{\om}{\omega}
\newcommand{\mfa}{\mathfrak{A}}

\newcommand{\mco}{\mathcal{O}}

\newcommand{\eps}{\varepsilon}
\newcommand{\fr}[2]{\frac{#1}{#2}}
\newcommand{\al}{\alpha}
\newcommand{\be}{\beta}

\newcommand{\real}{\mathbb{R}}

\newcommand{\ov}{\overline}

\newcommand{\non}{\nonumber}

\def\proof{\noindent{\bf Proof. }}
\def\qed{$\Box$\medskip}

\newtheorem{theoreme}{Theorem } [section]
\newtheorem{proposition}[theoreme]{Proposition}
\newtheorem{lemma}[theoreme]{Lemma}
\newtheorem{definition}[theoreme]{Definition}
\newtheorem{corollary}[theoreme]{Corollary}
\newtheorem{example}[theoreme]{Example}
\newtheorem{criterion}[theoreme]{Criterion}
\theoremstyle{remark}
\newtheorem{remark}[theoreme]{Remark} 
\newcommand{\beq}{\begin{equation}}
\newcommand{\eeq}{\end{equation}}
\newcommand{\beqa}{\begin{eqnarray}}
\newcommand{\eeqa}{\end{eqnarray}}
\newcommand{\ben}{\begin{arabicenumerate}}
\newcommand{\een}{\end{arabicenumerate}}
\newcommand{\bex}{\begin{example}}
\newcommand{\eex}{\end{example}}
\newcommand{\ber}{\begin{remark}}
\newcommand{\eer}{\end{remark}}
\newcommand{\bec}{\begin{corollary}}
\newcommand{\eec}{\end{corollary}}
\newcommand{\bed}{\begin{definition}}
\newcommand{\eed}{\end{definition}}
\newcommand{\bep}{\begin{proposition}}
\newcommand{\eep}{\end{proposition}}
\newcommand{\becr}{\begin{criterion}}
\newcommand{\eecr}{\end{criterion}}
\def\bel{\begin{lemma}}
\def\eel{\end{lemma}}
\def\bet{\begin{theoreme}}
\def\eet{\end{theoreme}}
\def\bed{\begin{definition}}
\def\eed{\end{definition}}

\def\RR{{\mathbb R}}

\def\ti{\tilde}

\def\id{{\rm id}}

\def\de{\delta}

\def\om{\omega}
\def\Om{\Omega}

\def\A{{\cal A}}
\def\B{{\mfa}}

\def\O{{\cal O}}

\def\I{{\cal I}}
\def\H{{\cal H}}
\def\K{{\cal K}}
\def\S{{\cal S}}

\def\emptyset{\varnothing}

\def\diffs1{{\rm Diff}(S^1)}

\def\Mob{{\rm\textsf{M\"ob}}}

\def\S2{S^{1(2)}}

\def\<{\langle}
\def\>{\rangle}

\def\supp{\mathrm{supp}\,}

\def\bP{\boldsymbol P}  

\begin{document}

\title{Asymptotic completeness for infraparticles in two-dimensional conformal field theory}

\author{
{\bf Wojciech Dybalski\footnote{Supported in part by the DFG grant SP181/25 and by the grant no. 09-065927 "Mathematical Physics" of the Danish Council for Independent Research.}}\\
Zentrum Mathematik, Technische Universit\"at M\"unchen,\\
D-85747 Garching, Germany\\
and\\
Department of Mathematical Sciences, Aarhus University,\\
DK-8000 Aarhus C, Denmark\\
E-mail: {\tt dybalski@ma.tum.de}
\and
{\bf Yoh Tanimoto\footnote{Supported in part by the ERC Advanced Grant 227458
OACFT ``Operator Algebras and Conformal Field Theory''.}}\\
Institut f\"ur Theoretische Physik, Universit\"at G\"ottingen,\\
Friedrich-Hund-Platz 1, D-37077 G\"ottingen, Germany\\
E-mail: {\tt yoh.tanimoto@theorie.physik.uni-goettingen.de}}
\date{}




\date{ }
\maketitle

\begin{abstract} We formulate a new concept of asymptotic completeness for two-dimensional massless quantum field theories in the
spirit of the theory of particle weights. We show  that this concept is more general than the standard particle interpretation based on 
Buchholz' scattering theory of waves. In particular, it holds in any chiral conformal field theory in an irreducible product representation
and in any  completely rational conformal field theory. This class  contains theories of infraparticles to which the scattering theory of waves does not apply. 
\end{abstract}

\noindent{\small {\bf Keywords:} infraparticles, conformal field theory, asymptotic completeness.}

\section{Introduction}

The problem of asymptotic completeness in quantum field theory (QFT) has been a subject of active  research
over the last two decades, both on the relativistic \cite{Le08,DT10} and non-relativistic side \cite{Sp97,DG99,FGS04,DK11,GMR11}. However, all the results obtained so
far concern Wigner particles, i.e., excitations with a well-defined mass. The problem of a complete particle 
interpretation in the presence of  infraparticles, i.e., particles whose mass fluctuates due to the presence of other excitations, 
appears to be open to date in all the models considered in the literature. In the present Letter we formulate a natural notion
of asymptotic completeness for two-dimensional massless relativistic QFT which remains meaningful in the presence of infraparticles.
We verify that a large class of chiral conformal field theories, containing theories of infraparticles, satisfies this property.

Since the seminal work of  Schroer  \cite{Sch63}, infraparticles have remained a prominent topic in mathematical physics.
Their importance relies on the fact that all the electrically charged particles, including the electron, turn out to be infraparticles \cite{Bu86}.
In models of non-relativistic QED scattering states of one electron and photons were successfully constructed by 
Fr\"ohlich, Pizzo and Chen in \cite{CFP07}. In a more abstract framework of algebraic QFT  a complementary approach to
scattering of infraparticles was proposed by Buchholz, Porrmann and Stein \cite{BPS91}. This theory of \emph{particle weights}
\cite{Po04.1,Po04.2,Dy10} does not aim at  
scattering states, but rather provides an algorithm for a direct construction of (inclusive) collision cross-sections. Very recently the
theory of particle weights was applied to conformal field theories by the present authors \cite{DT11}. We found out that any chiral
conformal field theory in a charged, irreducible product representation describes infraparticles. We also checked that in some cases 
these infraparticles have superselected velocity, similarly to the electron in QED. However, the question of complete particle interpretation
of these theories was not addressed in \cite{DT11}. We answer this question (affirmatively)  in the present work.

This Letter is organized as follows:  In Section~\ref{framework} we specify our framework and formulate a generalized concept
of asymptotic completeness (Definition~\ref{Asymptotic-completeness}). We remark that this is an implementation of ideas from 
\cite{Bu87} in the setting of  two-dimensional  massless theories.  
In Section~\ref{waves} we recall from \cite{Bu75} the scattering theory of waves, which are  counterparts of Wigner particles in this setting.   We show that any theory which has complete particle interpretation in the sense of waves satisfies also our generalized property of asymptotic completeness.
After this consistency check, we show in Section~\ref{infraparticles} that any chiral conformal field theory in an irreducible product representation
is asymptotically complete in the generalized sense. 
As a corollary, we obtain in Section~\ref{rational} that any completely rational conformal field theory has the property
of generalized asymptotic completeness.

\vspace{0.5cm}

\noindent\bf Acknowledgements. \rm
W.D.\! would like to thank D.\! Buchholz,  J.S.\! M\o ller, A.\! Pizzo, M.\!~Porrmann, 
W.\! De Roeck and H.\! Spohn for interesting discussions on scattering theory.
A part of this work has been accomplished during the stay of Y.T.  at Aarhus University.
He thanks J.S.\! M\o ller for his hospitality.

\section{The generalized concept of asymptotic completeness} \label{framework}
In this section we fix our framework, list the main definitions and facts relevant to
our investigation and formulate the generalized concept of asymptotic completeness.
We start with  a variant of the Haag-Kastler postulates \cite{Ha} which we will use in this work:
\begin{definition}\label{two-dim-net}
A local net of von Neumann algebras on $\real^2$ is a pair $(\B,U)$ consisting of 
a map $\mco\mapsto \B(\mco)$ from the family of open, bounded regions of $\real^2$ to the family of von Neumann algebras on a Hilbert space 
$\hil$, and a strongly continuous  unitary representation of translations $\real^2\ni \x\mapsto U(\x)$ acting on $\hil$, 
which are subject to the following conditions:
\begin{enumerate}
\item (isotony) If $\mco_1 \subset \mco_2$, then $\B(\mco_1)\subset \B(\mco_2)$.
\item (locality) If $\mco_1 \perp \mco_2$, then $[\B(\mco_1),\B(\mco_2)] = 0$, where $\perp$ denotes
spacelike separation.
\item (covariance)  $U(\x)\B(\mco)U(\x)^*=\B(\mco+\x)$ for any $\x\in\real^2$.
\item (positivity of energy) The joint spectrum of  $U$ coincides with the closed forward lightcone~$V_+:=\{\,(\om,\pb)\in\real^2\,|\,\om\geq |\pb|\,\}$. 
\end{enumerate}
We also introduce the quasilocal $C^*$-algebra of this net  $\B=\ov{\bigcup_{\mco\subset\real^2}\B(\mco)}$.
\end{definition}

We assume that the spectrum of $U$ coincides with $V_+$ rather than being included in,
because we are interested in the scattering theory of massless particles. It is indeed automatic
for dilation-covariant theories or theories of waves (see Section \ref{waves}).

Our first task is to identify, in the above theoretical setting, observables which can be interpreted as
particle detectors. To this end, we have to list several definitions and results:
First, we recall that an observable $B\in \B$ is called almost-local, 
if there exists a net of operators $\{\, B_r\in \B(\mco_r)\,|\, r>0\,\}$, s.t.\! for any $k\in\nat_0$
\beq
\lim_{r\to\infty} r^k\|B-B_r\|=0,
\eeq
where $\mco_r=\{(t,\xb)\in\real^2\,|\, |t|+|\xb|<r\,\}$. 
We also recall that the Arveson spectrum of an operator $B\in \mfa$ w.r.t.\! the group of translation automorphisms 
$\al_{\x}(\,\cdot\,)=U(\x)\,\cdot\,U(\x)^*$, denoted by $\Sp^B \al$,
is the closure of the union of supports of the distributions
\beq
(\Psi_1|\ti B(p)\Psi_2)=(2\pi)^{-1}\int_{\real^2} d x\,\e^{-ip\x} (\Psi_1|B(\x)\Psi_2) \label{energy-momentum-transfer}
\eeq
over all $\Psi_1,\Psi_2\in\hil$, where $p=(\om,\pb)$, $\x=(t,\xb)$, $p\x=\om t-\pb\xb$ and $B(\x):=\al_{\x}(B)$. 
Let $\PP(\,\cdot\,)$ be the spectral measure of $U$. As shown in \cite{Ar82},  for any $B\in\B$ and any closed set $\De\subset\real^2$,
it holds that
\beqa
B\PP(\De)\hil\subset \PP(\ov{\De+\Sp^B \al})\hil. \label{Arveson}
\eeqa 
Next, we introduce the lightline coordinates $\vv_{\pm}=\fr{\om\pm \pb}{\sqrt{2}}$, $t_{\pm}=\fr{t\mp \xb}{\sqrt{2}}$ 
and define, for any $\de>0$, the following subspaces of $\mfa$: 
\beqa
\mathcal L_{\pm,\de}=\{\, B\in\mfa\,|\, B\,\, \textrm{is almost-local and } \Sp^B\al\subset \{\, \vv_\pm\leq -\de \,\} \textrm{ is compact }\}.
\eeqa
Following \cite{AH67,Bu90}, we construct particle detectors: For any $e_{\pm}>0$,   $B_{\pm}\in \mathcal L_{\pm, e_{\pm}}$, $T\geq 1$
and $0<\eta<1$ we define
\beqa
Q^{T,\eta}_{\pm}(B_{\pm})=\int dt\,h_T(t) \int\, d\xb\, f_{\pm}^{\eta}(\xb/t)  (B^*_{\pm}B_{\pm})(t,\xb), \label{as-obs-one}
\eeqa
where $h_T(t)=|T|^{-\eps}h(|T|^{-\eps}(t-T))$, $0<\eps<1$ and $h\in C_0^{\infty}(\real)$ is a non-negative function
s.t.\! $\int dt\, h(t)=1$ and $f_{\pm}^{\eta}\in C^{\infty}(\real)$ have the following properties: $0\leq  f_{\pm}^\eta \leq 1$,
$f_{+}^{\eta}(\xb)=1$ for $\xb\geq \eta$, $f_{+}^{\eta}(\xb)=0$ for $\xb\leq 0$, $f_-^{\eta}(\xb)=f_+^{\eta}(-\xb)$. Moreover,
$f_{\pm}^{\eta}(\xb)\nearrow \mathbf 1_{\real_{\pm}}(\xb)$ as $\eta\to 0$, where $\mathbf 1_{\real_{\pm}}$ are the characteristic functions of the sets $\real_{\pm}$. For large positive\footnote{In the present Letter we consider only outgoing configurations of
particles, since the incoming case is analogous.  }  $T$ and small $\eta$
these expressions can be interpreted as  detectors sensitive to right-moving (in the $(+)$ case) and left-moving (in the $(-)$ case) particles. 


The operators $Q^{T,\eta}_{\pm}(B_{\pm})$ are defined on the domain $\mathcal D=\bigcup_{n\in\nat} \PP(\{(\om,p)\,|\, \om\leq n\})\hil$
of vectors of bounded energy. This is a consequence of  the following
abstract theorem due to Buchholz,  which we will use frequently in this paper:
\bet[\cite{Bu90}]\label{H-A} Let $\real^s\ni \xb\mapsto U(\xb)$ be a  group of unitaries on $\hil$,  $B\in B(\hil)$, $n\in \nat$ and let $\PP_n$ be the orthogonal projection onto the intersection of the kernels of the $n$-fold products $B(\xb_1)...B(\xb_n)$ for arbitrary $\xb_1,\ldots,\xb_n\in\real^{s}$, where 
$B(\xb)=U(\xb)BU(\xb)^*$. 
 Then there holds for each compact subset $K\subset \real^s$ the estimate
\beq
\left\|\PP_n\int_K d\xb\,(B^*B)(\xb)\PP_n\right\|\leq (n-1)\int_{\De K} d\xb\, \left\|[B^*,B(\xb)]\right\|,
\label{harmonic-analysis-bound}
\eeq
where $\De K=\{\, \xb-\yb\,|\, \xb,\yb\in K\,\}$.
\eet
\noindent As noticed in \cite{Bu90}, if $B\in\mathcal L_{\Ri,\de}\cup \mathcal L_{\Le,\de}$, then, for any compact set $\De$, the range of $\PP(\De)$ is contained in $\PP_n$ for sufficiently large $n$ due to relation (\ref{Arveson}). Exploiting almost-locality of $B$ one can replace  $\De K$ on the
r.h.s.\! of (\ref{harmonic-analysis-bound}) with $\real$, obtaining a bound which is uniform in $K$. Then $\PP(\De)\int d\xb\,(B^*B)(\xb)\PP(\De)\in B(\hil)$
exists as a strong limit of integrals over compact subsets and 
\beq
\int d\xb\, (B^*B)(\xb)\PP(\De)\in B(\hil), \label{harmonic-analysis}
\eeq
since $\Sp^B\al$ is compact.

 The existence of the limits $Q^{\tout,\eta}_{\pm}\Psi:= \lim_{T\to\infty}Q^{T,\eta}_{\pm}(B_{\pm})\Psi$, $\Psi\in \mathcal D$, is not known in general.
 If they exist for any $\Psi \in \mathcal D$, they define operators $Q^{\tout,\eta}_{\pm}(B_{\pm})$ on $\mathcal D$.
We show that these operators are translation-invariant in Lemma \ref{lm:translation-invariance} (cf. Proposition~3.9 of \cite{Po04.1}):
\beq
U(x)Q^{\tout,\eta}_{\pm}(B_{\pm})U(x)^*=Q^{\tout,\eta}_{\pm}(B_{\pm}),\quad x\in\real^2. \label{translation-invariance}
\eeq
In particular, they  preserve each spectral subspace of $U$. By the properties 
of functions $f^{\eta}_{\pm}$ and (\ref{harmonic-analysis}),  $\{Q^{\tout,\eta}_{\pm}(B_{\pm}) \}_{\eta\in (0,1)}$ are monotonously 
increasing (as $\eta\to 0$)
families of bounded operators on $\hil(\De):=E(\De)\hil$ which are uniformly bounded. Thus there exist the limits $Q^\tout_{\pm}(B_{\pm})
=\slim_{\eta\to 0} Q^{\tout,\eta}_{\pm}(B_{\pm} )$ as bounded operators on $\hil(\De)$. Since $\De$ is an arbitrary compact set, $Q^\tout_{\pm}(B_{\pm})$ can be consistently defined as  operators on $\mathcal D$, which also satisfy (\ref{translation-invariance}). 
Keeping the above discussion in mind, we define the following subsets of  $\mathcal L_{\pm,\de}$:
\begin{align}
\hat{\mathcal  L}_{\pm,\de}=&\{\,B_{\pm}\in \mathcal L_{\pm,\de}\,| \, Q^\tout_{\pm}(B_{\pm})\Psi:=\lim_{\eta\to 0}\lim_{T\to\infty}Q^{T,\eta}_{\pm}(B_{\pm})\Psi\,
\textrm{ exists for any } \Psi\in \mathcal D  \}. 
\end{align}
Every vector from the range of  $Q^\tout_{\Ri}(B_{\Ri})$ 
(resp. $Q^\tout_{\Le}(B_{\Le})$) contains
an excitation moving to the right (resp. to the left), whose energy is larger than $e_{\Ri}/\sqrt{2}$ 
(resp. $e_{\Le}/\sqrt{2}$), and possibly some other, unspecified excitations. The basis for this physical interpretation 
of particle detectors is Proposition~\ref{detectors-on-waves}, stated  below. 

Now, for any $\eps>0$, we define the following subset of the  spectrum of $U$
\beqa
\De_{\eps}(e_\Ri,e_{\Le})=\{\,(\vv_+,\vv_-)\in\real^2\,|\, e_{\Ri}\leq \vv_+\leq e_{\Ri}+\eps, \,  e_{\Le}\leq \vv_-\leq e_{\Le}+\eps\,\}.
\eeqa
Let $\hil_{\cc}$ be the continuous subspace of the relativistic mass operator $H^2-\Pb^2$, where $(H,\Pb)$ are the generators
of $U$. 
Then, in view of the above discussion, every non-zero vector of the form
\beqa
\Psi^\tout_{\eps}= Q^\tout_{\Ri}(B_{\Ri})Q^\tout_{\Le}(B_{\Le})\Psi,\quad \Psi\in \PP(\De_{\eps}(e_\Ri,e_{\Le}))\hil_{\cc},\quad B_{\pm}\in \hat{\mathcal  L}_{\pm,\de} \label{scattering-states}
\eeqa
describes two `hard' massless excitations, the first moving to the right with energy  $e_\Ri/\sqrt{2}$ and the
second moving to the left with energy $e_\Le/\sqrt{2}$, as well as some unspecified `soft' massless particles,
whose total energy is less than $\sqrt{2}\eps$.
Since the motion of massless excitations in two-dimensional Minkowski spacetime
is dispersionless, we expect that such two-body generalized scattering states span the entire subspace $\hil_{\cc}$. (In fact, two
excitations moving without dispersion in the same direction can be interpreted as one excitation). In view of
the above discussion, we define the generalized asymptotic completeness as follows:
\bed\label{Asymptotic-completeness} Suppose that for any $e_{\Ri},e_{\Le},\eps>0$ 
\beq
 \PP(\De_{\eps}(e_\Ri,e_{\Le}))\hil_{\cc}=
\Span\{\,Q^\tout_{\Ri}(B_{\Ri})Q^\tout_{\Le}(B_{\Le})\PP(\De_{\eps}(e_\Ri,e_{\Le})) \hil_{\cc} \,|\,  B_{\pm}\in \hat{\mathcal L}_{\pm, e_{\pm} } \, \}^{\cl},
\eeq
where ${\cl}$ means the closure.
Then we say that the theory has the property of generalized asymptotic completeness.
\eed
In the present Letter we show that this property is a generalization of a more standard concept
of asymptotic completeness in the sense of waves (Section~\ref{waves}). We provide a large class
of examples   which are not asymptotically complete in the sense of waves, but have the
generalized particle interpretation in the sense of Definition~\ref{Asymptotic-completeness} 
(e.g. charged sectors of chiral conformal field theories). However, we do not expect that the
generalized asymptotic completeness holds in all theories satisfying the postulates from Definition~\ref{two-dim-net}.
It may fail in models with too many local degrees of freedom, as for example certain generalized free fields.
We refrain from giving concrete counterexamples here.
 

\section{Theories of waves}\label{waves}
\setcounter{equation}{0}
In this section we consider a local net of von Neumann algebras $(\B,U)$ in a vacuum representation. That is
we assume, in addition to the properties specified in Definition~\ref{two-dim-net}, the existence of
a unique (up to a phase) unit vector $\Om\in\hil$, which is invariant under $U$ and cyclic for $\mfa$. Let
$\hil_{\pm}=\ker(H\mp \bP)$, where $(H,\bP)$ are generators of $U$, and let $\PP_{\pm}$ be the corresponding orthogonal projections.
If each of the subspaces $\hil_{\pm}$ contains some vectors orthogonal
to $\Om$, then we say that the net $(\B,U)$ describes `waves', which are  counterparts of Wigner particles in massless,
two-dimensional theories. A natural scattering theory for waves, developed by Buchholz in \cite{Bu75}, is outlined below.
 We will show that theories which are asymptotically complete in the sense of this scattering theory have also the
property of generalized asymptotic completeness, formulated in Definition~\ref{Asymptotic-completeness} above.

Following \cite{Bu75}, for any $F\in\B$ and $T\geq 1$ we introduce the asymptotic field approximants:
\beqa
F_\pm(h_T)= \int h_T(t) F(t, \pm t) dt,
\eeqa
where $h_T$ is defined after formula~(\ref{as-obs-one}) above. 
We recall the following result:
\bep[\cite{Bu75}]\label{scattering-first-lemma} Let $F\in \B$.  Then the limits
\beqa
\Phi_{\pm}^{\tout}(F):= \underset{T\to\infty}\slim \, F_{\pm}(h_T) \quad \label{asymptotic-field}
\eeqa
exist and are called the (outgoing) asymptotic fields. They depend only on the respective vectors $\Phi_{\pm}^{\tout}(F)\Om=\PP_{\pm}F\Om$ 
and satisfy $[\Phi_{+}^{\tout}(F),\Phi_{-}^{\tout}(F')]=0$ for any $F,F'\in\mfa$.
\eep
 Now the scattering states are defined  as follows: Since $\B$ acts irreducibly
on $\hil$ (by the assumed uniqueness of the vacuum), for any $\Psi_{\pm}\in\hil_{\pm}$ we can find $F_{\pm}\in\B$ s.t.\! $\Psi_{\pm}=F_{\pm}\Om$ \cite{Sa}. The vectors
\beqa
\Psi_+\pout\Psi_-=\Phi_{+}^{\tout}(F_+)\Phi_{-}^{\tout}(F_-)\Om \label{wave-scattering-states}
\eeqa
are called the (outgoing) scattering states. By Proposition~\ref{scattering-first-lemma} they do not depend on the choice of
$F_{\pm}$ within the above restrictions. They have the following properties: 
\bep[\cite{Bu75}]\label{scattering-second-lemma} Let $\Psi_{\pm},\Psi_{\pm}'\in\hil_{\pm}$. Then:
\begin{enumerate}
\item[(a)] $(\Psi_+\pout \Psi_-,\Psi'_+\pout \Psi'_-)=(\Psi_+,\Psi'_+)(\Psi_-,\Psi'_-)$,
\item[(b)] $U(\x)(\Psi_+\pout \Psi_-)=(U(\x)\Psi_+)\pout (U(\x)\Psi_-)$, for $\x\in\real^2$.
\end{enumerate}
\eep
If the states of the form~(\ref{wave-scattering-states}) span the entire Hilbert space, then we say that the theory
is {\bf asymptotically complete in the sense of waves}. In this case, the representation $U$
decomposes into a tensor product of representations of lightlike translations and
the spectrum of $U$ automatically coincides with $V_+$ by the theorem of Borchers \cite{Bo92}.
We will show below that any such theory is also asymptotically
complete in the sense of Definition~\ref{Asymptotic-completeness}. To this end we prove the following fact:
\bep\label{detectors-on-waves} Let $B_{\pm}\in \mathcal L_{\pm, e_{\pm}}$ and let $\Psi_{\pm}\in\hil_{\pm}$ be vectors of bounded 
energy. Then
\beqa
\lim_{\eta\to 0}\lim_{T\to\infty}Q^{T,\eta}_{\Ri}(B_{\Ri})(\Psi_+\pout \Psi_-)&=&(\PP_+Q(B_{\Ri})\Psi_+)\pout \Psi_-, \label{convergence-one}\\  
\lim_{\eta\to 0}\lim_{T\to\infty}Q^{T,\eta}_{\Le}(B_{\Le}) (\Psi_+\pout \Psi_-)&=&\Psi_+\pout (\PP_-Q(B_{\Le})\Psi_-),\label{convergence-two}
\eeqa
where $Q(B_{\pm}):=\int d\xb\,(B_{\pm}^*B_{\pm})(\xb)$ are operators defined on $\mathcal D$.
\eep
\proof We prove only  equality (\ref{convergence-one}), as (\ref{convergence-two}) is analogous. Let $F_{\pm}\in\mfa$
be s.t.\! $\Psi_{\pm}=F_{\pm}\Om$. Since $\Psi_{\pm}$ have bounded energy, we can ensure, by smearing with suitable
test functions, that $\Sp^{F_{\pm}}\al$ are compact sets.  Then it is clear that $\Psi_+\pout \Psi_-=\Phip(F_+)\Phim(F_-)\Om$
is a vector of bounded energy, and we can write
\beqa
Q^{T,\eta}_{\Ri}(B_{\Ri}) \Phip(F_+)\Phim(F_-)\Om&=&[Q^{T,\eta}_{\Ri}(B_{\Ri}), \Phim(F_-)]\Phip(F_+)\Om\non\\
& &+\Phim(F_-)Q^{T,\eta}_{\Ri}(B_{\Ri})\Phip(F_+)\Om. \label{commuting-detectors}
\eeqa
Let us first consider  the second term on the r.h.s.\! of (\ref{commuting-detectors}). 
We define
\beqa
Q^T_{\Ri}(B_{\Ri}):=\int dt\,h_T(t) \int\, d\xb\, \mathbf{1}_{\real_+} (\xb/t)  (B_+^*B_+)(t,\xb). 
\eeqa
We note that $R_-^{T,\eta}(B_+):=Q^T_{\Ri}(B_{\Ri})-Q^{T,\eta}_{\Ri}(B_{\Ri})$ satisfies the
assumptions of Lemma~\ref{technical-commutators}. Consequently,
\beqa
\lim_{T\to\infty} R_-^{T,\eta}(B_+)\Phip(F_+)\Om=
\lim_{T\to\infty} [R_-^{T,\eta}(B_+), F_+(h_T)]\Om=0, \label{new-formula}
\eeqa
where we made use of the fact that $\sup_{T\in \real}\| R_-^{T,\eta}(B_+)E(\De)\|<\infty$
for any compact set $\De$. Now we compute
\beqa
Q^T_{\Ri}(B_{\Ri})\Psi_+&=&\int dt\,h_T(t) \e^{iHt}\int_{\real_+}\, d\xb (B^*_{+}B_{+})(\xb)\e^{-i\Pb t}\Psi_+\non\\
&=&\int dt\,h_T(t) \e^{i(H-\Pb)t}\int_{-t}^{\infty}\, d\xb (B^*_{+}B_{+})(\xb)\Psi_+\non\\
&=&-\int dt\,h_T(t) \e^{i(H-\Pb)t}\int_{-\infty}^{-t}\, d\xb (B^*_{+}B_{+})(\xb)\Psi_+\non\\ 
& &+\left(\int dt\,h_T(t) \e^{i(H-\Pb)t}-\PP_+\right)\int\, d\xb (B^*_{+}B_{+})(\xb)\Psi_+\non\\
& &+\PP_+\int\, d\xb (B^*_{+}B_{+})(\xb)\Psi_+.   \label{detector-on-sp-state}
\eeqa
Here in the first step we made use of the definition of $\hil_+$.   The second term on the r.h.s. above 
tends to zero as $T\to \infty$ by the mean ergodic theorem.
Let us show that the  first term on the r.h.s. of (\ref{detector-on-sp-state}) tends to zero as $T\to\infty$. This is a consequence
of the fact that
\beq
\lim_{t\to\infty}\int_{-\infty}^{-t}\, d\xb (B^*_{+}B_{+})(\xb)\Psi_+=0
\eeq
which follows from the discussion after Theorem~\ref{H-A} above. Thus we obtain 
\beq
\lim_{T\to\infty}Q^T_{\Ri}(B_{\Ri})\Psi_+=\PP_+Q(B_+)\Psi_+. \label{second-term}
\eeq
To conclude the proof, we still have to show that the first term on the r.h.s.\! of (\ref{commuting-detectors}) tends strongly to
zero as $T\to\infty$. This follows from the equality
\beq
\lim_{T\to\infty}\|\PP(\De)[Q^{T,\eta}_{\Ri}(B_{\Ri}), F_-(h_T)]\PP(\De')\|=0, \label{commutator-vanishing}
\eeq
valid for any compact sets $\De,\De'\subset V_+$, which is established in  Lemma~\ref{technical-commutators}. In fact, let us consider separately the two terms forming the commutator
in (\ref{commuting-detectors}):
\begin{align}
Q^{T,\eta}_{\Ri}(B_{\Ri})\Phim(F_-)\Phip(F_+)\Om&=Q^{T,\eta}_{\Ri}(B_{\Ri}) F_-(h_T)\Phip(F_+)\Om+o(1),\label{first-term-comm}\\
\Phim(F_-)Q^{T,\eta}_{\Ri}(B_{\Ri})\Phip(F_+)\Om&=\Phim(F_-)\PP_+Q(B_+)  \Phip(F_+)\Om+o(1)\non \\
&=F_-(h_T)\PP_+Q(B_+)  \Phip(F_+)\Om+o(1)\non\\
&=F_-(h_T)Q^{T,\eta}_{\Ri}(B_{\Ri})\Phip(F_+)\Om+o(1), \label{second-term-comm}
\end{align}
where $o(1)$ denotes terms tending in norm to zero as $T\to\infty$. In (\ref{first-term-comm}) we used the fact that  
$\Sp^{F_{\pm}}\al$ are compact and relation (\ref{harmonic-analysis}) which gives $\sup_{T\in\real }\|Q^{T,\eta}_{\Ri}(B_{\Ri})E(\De)\|<\infty$
for any compact set $\De$. In the first and last step of (\ref{second-term-comm}) we exploited (\ref{second-term}) and (\ref{new-formula}).
>From (\ref{first-term-comm}), (\ref{second-term-comm}) and the compactness of $\Sp^{B_{\Ri}}\al$ we conclude that
(\ref{commutator-vanishing}) implies vanishing of the  first term on the r.h.s.\! of (\ref{commuting-detectors}) as $T\to\infty$. \qed

Let us set $H_{\pm}=\fr{1}{\sqrt 2}(H\pm \Pb)$ and let $\PP_{\pm}(\,\cdot\,)$ be the spectral measures of $H_{\pm}|_{\hil_{\pm}}$. 
It is easily seen that the spectrum of $H_{\pm}|_{\hil_{\pm}}$ is continuous, apart from an eigenvalue at zero. (In fact, if $\Psi_+$
is an eigenvector of $H_+|_{\hil_+}$, then  $(\Psi_+|A\Psi_+)=(e^{itH_+}\Psi_+|Ae^{itH_+}\Psi_+)=(e^{it\sqrt 2 \Pb}\Psi_+|Ae^{it\sqrt 2 \Pb}\Psi_+)=(\Psi_+|A(0,-\sqrt 2 t)\Psi_+)=\|\Psi_+\|^2(\Om|A\Om)$
for any $A\in\mfa$ by the clustering property. Exploiting the fact that $\mfa$ acts irreducibly on $\hil$, we obtain that $\Psi_+$ is proportional to $\Om$). 
We note the following fact, whose proof relies on some ideas from the proof of Proposition~2.1 of \cite{BF82}:
\begin{lemma}\label{borchers-zero}
Let $\de>0$, $\Psi_+\in \hil_+$ and  suppose that $B\Psi_+=0$ for 
any $B\in\mathcal L_{+,\de}$. Then $\PP_+([\de,\infty))\Psi_+=0$. (An analogous result holds for $(+)$ replaced with $(-)$). 
\end{lemma}
\proof Let us choose $b>a>\de$, $0<\eps<a-\de$ and $c>0$. We choose  functions $f_{\pm}\in S(\real)$ s.t.\! $\supp \ti f_+\subset (-\infty, -\de  ]$ is compact, $\ti f_+(\vv_+)=1$ for $\vv_+\in [-b,-a+\eps]$, $\supp\,\ti f_-\subset [-2c,2c]$ and $\ti f_-(\om_-)=1$ for $\om_-\in [-c,c]$. Let $f(x)=f_+(t_+)f_-(t_-)$. 
Since $\ti f(p)=\ti f_+(\vv_+)\ti f_-(\vv_-)$,  we obtain that $A(f)=\int_{\real^2} dx\,A(x)f(x)$ is an element of  $\mathcal L_{+,\de}$ for any $A\in\mfa(\mco)$, $\mco\subset\real^2$. Thus, by assumption,
$A(f)\Psi_+=0$. 
Making use of the fact that $\al_x(A(f))\in \mathcal L_{+,\de}$ for any $x\in\real^2$,
we obtain that $U(x)A(f)U(x)^*\Psi_+=0$, hence $A(f)U(x)^*\Psi_+=0$ and consequently
\beq
\PP(\De_2)A(f)\PP(\De_1)\Psi_+=0,
\eeq    
for any compact sets $\De_1$, $\De_2\subset \real^2$. Setting $\De_1=\{\,(\vv_+,\vv_-)\in\real^2\,|\,\vv_+\in [a,b],\vv_-\in [-c/2,c/2]\,\}$, 
$\De_2=\{\, (\vv_+,\vv_-)\in\real^2      \,|\, \vv_+\in [0,\eps], \vv_-\in [-c/2,c/2]   \,\}$ and exploiting the properties
of  $f$, we obtain that
\beq
\PP(\De_2)A\PP(\De_1)\Psi_+=0.
\eeq 
As $\mfa$ acts irreducibly on $\hil$, (since we assumed the uniqueness of the vacuum vector), and $\PP(\De_2)\neq 0$, (which follows e.g.\! from the existence of the vacuum), we conclude that $\PP(\De_1)\Psi_+=\PP_+([a,b])\Psi_+=0$. Since the spectrum of $H_+|_{\hil_+}$ is 
continuous, apart from the eigenvalue at zero, we obtain that $\PP_+([\de,\infty))\Psi_+=0$. \qed\\
Now we proceed to the main result of this section:
\bet\label{main-theorem-waves} Let $(\B,U)$ be a net of von Neumann algebras  in a vacuum representation, which is asymptotically
complete in the sense of waves. Then it has the property of generalized asymptotic
completeness, stated in Definition~\ref{Asymptotic-completeness}.
\eet
\proof First, we note that the continuous subspace $\hil_{\cc}$ of the relativistic mass operator $H^2-\Pb^2$ is given by 
\beq
\hil_{\cc}=\hil_{+,\cc}\pout\hil_{-,\cc}, 
\label{continuous-spectrum}
\eeq
where $\hil_{\pm,\cc}=\hil_{\pm}\cap\{\Om\}^{\bot}$ are the continuous subspaces of $H_{\pm}|_{\hil_{\pm}}$. 
To justify this fact one notes that, as a consequence of 
asymptotic completeness in the sense of waves, $\hil_{\cc}\subset \hil_{+,\cc}\pout\hil_{-,\cc}$ and the only possible eigenvalue of $H^2-\Pb^2$ is zero. 
(Non-zero eigenvalues can easily be excluded  with the help of the Haag-Ruelle scattering theory or by proceeding as in 
Lemma~\ref{continuous-spectra} below). 
Then it is readily checked that no vector from the subspace on the r.h.s.\! of~(\ref{continuous-spectrum}) can be a corresponding eigenvector.

Making use of (\ref{continuous-spectrum}) and of Proposition~\ref{scattering-second-lemma}, we obtain  the following equality
\beqa
\PP(\De_{\eps}(e_\Ri,e_{\Le}))\hil_{\cc}=\PP_+([e_{\Ri},e_{\Ri}+\eps] )\hil_{+,\cc}\pout \PP_-([e_{\Le},e_{\Le}+\eps] )\hil_{-,\cc}.
\label{tensor-product-measures}
\eeqa
Now we note that any vector $\Psi\in \PP(\De)\hil$, where $\De\subset \real^2$ is compact,  can be expressed as 
$\Psi=\sum_{m,n}c_{m,n}\Psi_{+,m}\pout \Psi_{-,n}$, where $\Psi_{\pm,m}\in P(\De')\hil_\pm$ 
form orthonormal systems, $\De'\subset \real^2$ is compact and $\sum_{m,n}|c_{m,n}|^2<\infty$. (See \cite[Lemma A.2]{DT11}).
Hence, Proposition~\ref{detectors-on-waves} and relation~(\ref{harmonic-analysis})  entail that for any $B_{\pm}\in\mathcal L_{\pm,e_{\pm}}$ and
$\Psi\in \mathcal D$ the limits $Q_{\pm}^\tout(B_\pm)\Psi=\lim_{\eta\to 0}\lim_{T\to\infty}Q^{T,\eta}_{\pm}(B_{\pm})\Psi$ exist. Consequently,  it suffices to verify the following formula
\beq
\PP_+([e_{\Ri},e_{\Ri}+\eps] )\hil_{+,\cc}=\Span\{\,\PP_+Q(B_+)\PP_+([e_{\Ri},e_{\Ri}+\eps] )\hil_{+,\cc} \,|\,  B_{+}\in \mathcal L_{+,e_{+}}  \}^{\cl} \label{one-dim-zero}
\eeq 
and its counterpart with $(+)$ replaced with $(-)$, whose proof is analogous. 
(We recall that  $Q(B_+)$ was defined in Proposition~\ref{detectors-on-waves}).
Since $\PP_+Q(B_+)\PP_+$ is invariant under spacetime translations, 
it is obvious that the subspace on the r.h.s.\! 
of (\ref{one-dim-zero}) is
contained in the subspace on the l.h.s. Let us now assume that the inclusion is proper i.e., we can
choose a non-zero vector $\Psi_+ \in (\PP_1-\PP_0)\hil_{+,\cc}$, where $\PP_1:=\PP_+([e_{\Ri},e_{\Ri}+\eps] )$ and $\PP_0$ is
the orthogonal projection on the subspace on the r.h.s.\! of (\ref{one-dim-zero}).
By Lemma~\ref{borchers-zero}, there is an operator $B_+\in\mathcal L_{+,e_+}$ 
s.t.\! $B_+\Psi_+ \ne 0$. Then it is easy to see that $(\Psi_+|Q(B_+)\Psi_+) \ne 0$
which means that $(\PP_1-\PP_0)\PP_+Q(B_+) \Psi_+\ne 0$. Hence  $\PP_+Q(B_+)\Psi_+ \neq 0$ and $\PP_+Q(B_+)\Psi_+\notin \PP_0\hil_{+,\cc}$,
which contradicts the definition of $\PP_0$. \qed

\section{Chiral nets and infraparticles}\label{infraparticles}
\setcounter{equation}{0}
In the previous section we showed that the generalized concept of particle interpretation, formulated
in Definition~\ref{Asymptotic-completeness}, is a consequence of a more standard notion of asymptotic
completeness in the sense of waves. We recall from \cite{DT10,DT11} that any chiral conformal field theory in a \emph{vacuum} product representation
is asymptotically complete in the sense of waves. Hence  it is also asymptotically complete in the generalized sense.

It turns out that the range of validity of the generalized asymptotic completeness is
not restricted to theories of waves, but includes also some theories of \emph{infraparticles}. We say that a net
of von Neumann algebras $(\mfa,U)$ describes infraparticles, if one or both of the subspaces $\hil_{\pm}$ are trivial,
but there exist non-zero particle detectors $Q_{\pm}^{\tout}(B_{\pm})$. As shown in \cite{DT11}, any chiral conformal
field theory in a \emph{charged} irreducible product representation describes infraparticles. In this section we show that these
theories of infraparticles are asymptotically complete in the generalized sense.

Let us now briefly recall the construction of chiral conformal field theories, focusing on these properties,
which are needed in our investigation. First, we recall the definition of a local net of von Neumann algebras on the  real line:  
\begin{definition}\label{circle-theory}
A local net of von Neumann algebras on $\real$ is a pair $(\A,\U)$ consisting of 
a map $\I\mapsto \A(\I)$ from the family of open, bounded subsets of $\real$ to the
family of von Neumann
algebras on a Hilbert space $\hilk$ 
and a strongly continuous  unitary representation of translations $\real\ni \t\mapsto \U(\t)$, acting on $\hilk$, 
which are subject to the following conditions:
\begin{enumerate}
\item (isotony) If $\I \subset \J$, then $\A(\I)\subset \A(\J)$.
\item (locality) If $\I \cap \J = \emptyset$, then $[\A(\I),\A(\J)] = 0$.
\item (covariance) $\be_{\t}(\A(\I)):=\U(\t)\A(\I)\U(\t)^* = \A(\I+\t)$ for any $\t\in\real$.
\item (positivity of energy) The spectrum of $\U$ coincides with $\real_+$. 
\end{enumerate}
We also denote by $\A$ the quasilocal $C^*$-algebra of this net i.e., $\A=\ov{\bigcup_{\I\subset\real}\A(\I)}$. We assume that 
it acts irreducibly on $\hilk$.
\end{definition}
Let $(\A_\l,\U_\l)$ and $(\A_\r,\U_\r)$ be two nets of von Neumann algebras on $\RR$,  acting on Hilbert spaces
$\hilk_\l$ and $\hilk_\r$. To construct a local net $(\B, U)$ on $\RR^2$, acting on the tensor product space $\H = \K_\l\otimes \K_\r$,  
we identify the two real lines with the lightlines $I_{\pm}=\{\,(t,\xb)\in\real^2\,|\, \xb\mp t=0\,\}$ in $\RR^2$. 
Let us  first specify the unitary representation of translations
\beq
U(t,\xb):= \U_\l\left(\fr{1}{\sqrt{2}}(t-\xb)\right)\otimes \U_\r\left(\fr{1}{\sqrt{2}}(t+\xb)\right). 
\label{unitary-representation}
\eeq
The spectrum of this representation coincides with $V_+$ due to property~4 from Definition~\ref{circle-theory}.  
Any double cone  $D\subset\RR^2$ can be expressed as a product of intervals
on lightlines $D = \I\times \J$. The corresponding local von Neumann algebra is given  by $\B(D):= \A_\l(\I)\otimes\A_\r(\J)$,
and for a general open region $\mco$ we put $\B(\mco)=\bigvee_{D\subset \mco}\B(D)$. The resulting  net of von Neumann algebras $(\B,U)$, 
which we call the chiral net, satisfies the properties stated in Definition~\ref{two-dim-net}. If both $\hilk_\l$ and $\hilk_\r$ contains
translation invariant vectors, then we say that the net  $(\B,U)$  is in a vacuum product representation. Otherwise we say that it is
in a charged product representation. These two cases will be treated on equal footing in the remaining part of this section. 

We will show that any chiral net satisfies the generalized asymptotic completeness in the sense of Definition~\ref{Asymptotic-completeness}. 
As a preparation, we prove the following two lemmas. 
\begin{lemma}\label{borchers} Let $(\A,\U)$ be a local net of von Neumann algebras on $\real$. 
\begin{enumerate}
\item[(a)] For any $\de>0$, we define the following subset of $\A$: 
\beq
\LLL_{\de}=\{\, A(f)\, |\, A\in\A(\I),  \I\subset \real, f\in S(\real), \supp\,\ti f\subset  (-\infty,-\de]\,\,\,\, \mathrm{ compact } \,\},
\label{LLL}
\eeq
where  $A(f):=\int d\t\,\be_{\t}(A)f(\t)$.
Suppose that $B\Psi=0$ for any $B\in \LLL_{\de}$. 
Then we have $\EE([\de,\infty))\Psi=0$, where $\EE(\,\cdot\,)$ is the spectral measure of $V$.
\item[(b)] The spectrum of $V$ is absolutely continuous, apart from a possible eigenvalue at zero. 
\end{enumerate}
\end{lemma}
\proof The argument below, which is a one-dimensional version of the proof of Lemma~\ref{borchers-zero},
 relies on ideas from  Proposition~2.1 and 2.2 of \cite{BF82}. To prove (a)
we  choose $b>a>\de$ and $0<\eps<a-\de$. We pick a function $f\in S(\real)$ s.t.\! $\supp \ti f\subset (-\infty, -\de  ]$ is compact 
and $\ti f(\om)=1$ for $\om\in [-b,-a+\eps]$. 
Then $A(f)\in \LLL_{\de}$ and, by assumption,  $A(f)\Psi=0$. Making use of the fact that $\be_s(A(f))\in \LLL_{\de}$ for any $s\in\real$,
we obtain
\beq
\EE(\De_2)A(f)\EE(\De_1)\Psi=0,
\eeq    
for any compact sets $\De_1$, $\De_2$. Setting $\De_1=[a,b]$, $\De_2=[0,\eps]$ and exploiting the properties
of the function $f$, we obtain that
\beq
\EE(\De_2)A\EE(\De_1)\Psi=0.
\eeq 
Since $\A$ acts irreducibly on $\hilk$, we conclude that $\EE([a,b])\Psi=0$. Thus we obtain that either $\EE([\de,\infty))\Psi=0$
or $V(s)\Psi=\e^{i\de s}\Psi$ for all $s\in\real$. 

Let us now exclude the latter possibility: By Lemma 2.2 of \cite{Bu90}, (stated as Theorem~\ref{H-A} above), we obtain that
\beq
\int ds\,(\Psi|\be_s(B^*B)\Psi)<\infty
\eeq
for any $B\in\LLL_{\de'}$, $\de'>0$.
This is only possible if $B\Psi=0$ for all such $B$. Proceeding as in the first part of the proof, we conclude that $V(s)\Psi=\Psi$ i.e., $\de=0$,
which is  a contradiction. This concludes the proof of (a).

To show (b), we pick $0<\eps<\de$ and note that 
\beq
\Span\{\, \EE([\de,\de+\eps]) B^*\Psi   \,|\, B\in \LLL_{\de},\, \Psi\in  \EE([0,\eps])\hilk \,\}^{\cl}=\EE([\de,\de+\eps])\hilk.
\eeq
In fact, any vector from $\EE([\de,\de+\eps])\hilk$, which is orthogonal to the subspace on the l.h.s.\! is zero
by relation~(\ref{Arveson}) and part (a) of the present lemma. Next, by irreducibility, for any $A(f)\in \LLL_{\de}$
where $\supp \tilde f \subset (-\infty, -\delta]$
we can find $A_n\in \A(\I_n)$ s.t. $\EE([\de,\de+\eps])A(f)^*=\slim_{n\to\infty} A_n(f)^*$. Consequently,
\beq
\Span\{\, B^*\Psi   \,|\, B\in \LLL_{\de},\, \Psi\in  \EE([0,\eps])\hilk \,\}^{\cl}\supset \EE([\de,\de+\eps])\hilk.
\eeq
Now for any $B_1, B_2\in \LLL_{\de}$ and $\Psi_1,\Psi_2\in  \EE([0,\eps])\hilk$, we get
\beqa
|(B_1^*\Psi_1|V(s)B_2^*\Psi_2)|&=&|(\Psi_1|[B_1, B_2^*(s)] V(s)\Psi_2)|\non\\
&\leq& \|\Psi_1\|\,\|\Psi_2\|\,\| [B_1, B_2^*(s)]\|,
\eeqa
which is a rapidly decreasing function of $s$. Making use of these facts and of the Plancherel theorem, one easily obtains that 
$(\Psi|\EE(\De)\Psi')=0$ for any $\Psi,\Psi'\in \EE((0,\infty))\hilk$ and $\De\subset \real$ of zero Lebesgue measure. \qed
\bel\label{continuous-spectra} Let $(\A_\l,\U_\l)$ and $(\A_\r,\U_\r)$ be two nets of von Neumann algebras on $\RR$ acting on $\hilk_\l$ and $\hilk_\r$, 
respectively, and let   $(\B,U)$ be the corresponding chiral net. Then
\beq
\hil_{\cc}=\hilk_{\l,\cc}\otimes\hilk_{\r,\cc}, \label{spectrum}
\eeq
where $\hil_{\cc}$ is the continuous subspace of $H^2-\Pb^2$ and $\hilk_{\l/\r,\cc}$ are the continuous subspaces
of $V_{\l/\r}$.
\eel
\proof Let $T_{\l/\r}$ be the generators of $V_{\l/\r}$ and $E_{L/R}$ their spectral measures. We obtain from relation~(\ref{unitary-representation}) that 
$\Pb=\fr{1}{\sqrt{2}}(T_\l\otimes I-I\otimes T_\r)$, 
$H=\fr{1}{\sqrt{2}}(T_\l\otimes I+I\otimes T_\r)$ and $H^2-\Pb^2=2(T_\l\otimes T_\r)$. Thus it follows immediately from
Lemma~\ref{borchers}~(b) that $\hil_{\cc}\subset\hilk_{\l,\cc}\otimes\hilk_{\r,\cc}$. To prove the opposite inclusion,
we have to show that the r.h.s.\! of (\ref{spectrum})  does not contain any eigenvectors of $H^2-\Pb^2$.  Let us therefore assume 
that there exists $\Psi\in  \hilk_{\l,\cc}\otimes\hilk_{\r,\cc}$ s.t.\! $(H^2-\Pb^2)\Psi=m^2\Psi$, $m\geq 0$. Then, for any $\Psi_{\l/\r}\in \hilk_{\l/\r,\cc}$,
we can write
\beqa
|(\Psi|\Psi_\l\otimes\Psi_\r)|
&=&|\int_{V_+} (\Psi| d E(q_\l,q_\r)(\Psi_\l\otimes \Psi_\r))|\non\\
&=&|\int_{H_m} (\Psi| d E(q_\l,q_\r)(\Psi_\l\otimes \Psi_\r))|\non\\
&\leq& \|\Psi\| (\Psi_\l\otimes \Psi_\r|  E(H_m)(\Psi_\l\otimes \Psi_\r))^\h, \label{spectral-measures}
\eeqa
where $d E( q_{\l},q_{\r})=dE_{\l}(q_{\l})\otimes dE_{\r}(q_{\r})$ is the joint spectral measure of $(H,\Pb)$ expressed in the lightcone coordinates
$q_{\l}:=\fr{\om+\pb}{\sqrt{2}}$, $q_{\r}:=\fr{\om-\pb}{\sqrt{2}}$. Here
$H_m=\{\,(q_{\l},q_{\r})\in\real_+^2\,|\,  q_{\l} q_{\r}=m^2/2\}$ is the hyperboloid at mass $m$ (or the boundary of the lightcone in the case $m=0$) and the second equality in (\ref{spectral-measures}) follows
from the assumption that $\Psi$ is an eigenvector of $H^2-\Pb^2$.  The measure $(\Psi_\l\otimes \Psi_\r|  E(\,\cdot\,)(\Psi_\l\otimes \Psi_\r))$
appearing in the last line of (\ref{spectral-measures}) 
is a product of Lebesgue absolutely continuous measures 
by  Lemma~\ref{borchers}~(b), hence it is also absolutely continuous.
Since $H_m$ has Lebesgue measure zero, the expression on the r.h.s. of (\ref{spectral-measures}) is zero.
Thus $\Psi=0$, which concludes the proof. \qed
\ber We note that the above lemma could be proven without exploiting the absolute continuity of the spectral measures. 
In fact,  for any two positive operators $T_\l, T_\r$ with empty  point spectrum,
the operator $T_\l\otimes T_\r$ also has empty point spectrum. This follows from the elementary fact
that if $\mu_\l,\mu_\r$ are two measures on $\RR$ without an atomic part, 
then the product measure $\mu_\l\times\mu_\r$ of any hyperboloid is zero.
\eer

\noindent Now we are ready to prove  our main result:
\bet Any chiral net $(\B,U)$ satisfies generalized asymptotic completeness in the sense of Definition~\ref{Asymptotic-completeness}.
\eet
\proof First, we obtain from Lemma~\ref{continuous-spectra},
\beq
\PP(\De_{\eps}(e_\Ri,e_\Le))\hil_{\cc}=\PP_\l([e_\Ri,e_\Ri+\eps])\hilk_{\l,\cc}\otimes \PP_\r([e_\Le,e_\Le+\eps])\hilk_{\r,\cc}, \label{spectral-measure}
\eeq
where $\hilk_{\l/\r,\cc}$ and $\PP_{\l/\r}(\,\cdot\,)$ are the continuous subspaces and spectral measures  of $V_{\l/\r}$. Now let $\LLL_{\l/\r,\de}\subset \A_{\l/\r}$ be sets defined
as in (\ref{LLL}). It is easy to see that 
if  $B_\l\in\LLL_{\l,e_{\Ri}}$ and $B_\r\in\LLL_{\r,e_{\Le}}$, then
$B_\l\otimes I\in \mathcal L_{\Ri,e_{\Ri}} $ and $I\otimes B_\r\in \mathcal L_{\Le,e_{\Le}}$. 
Moreover, we obtain
\beqa
\lim_{\eta\to 0}\lim_{T\to\infty}Q^{T,\eta}_{\Ri}(B_\l\otimes I)(\Psi_\l\otimes\Psi_\r)&=& (Q(B_\l)\Psi_\l)\otimes \Psi_\r,\label{detector-convergence-one}\\
\lim_{\eta\to 0}\lim_{T\to\infty}Q^{T,\eta}_{\Le}(I\otimes B_\r)(\Psi_\l\otimes\Psi_\r)&=&\Psi_\l\otimes (Q(B_\r)\Psi_\r), \label{detector-convergence-two}
\eeqa
where $Q(B_{\l/\r}):=\int d\t\, \be^{(\l/\r)}_{\t/\sqrt{2}}(B_{\l/\r}^*B_{\l/\r})$,
$\Psi_{\l/\r}\in \PP_{\l/\r}(\De_{\l/\r})\hilk_{\l/\r}$ and $\De_{\l/\r}\subset\real$ are compact subsets.
To verify relation~(\ref{detector-convergence-one}), we note that
\beqa
& &\lim_{T\to\infty}Q^{T,\eta}_{\Ri}(B_\l\otimes I)(\Psi_\l\otimes\Psi_\r)\non\\
& &\ph{44444444}=\lim_{T\to\infty} \left(\int dt\,h_T(t)\int_{-\infty}^{\infty} ds\, f_+^{\eta}(1-s/t)  
\be^{(\l)}_{\t/\sqrt{2}}(B_{\l}^*B_{\l})\Psi_\l\right)\otimes \Psi_\r
\non\\
& &\ph{44444444}=(Q(B_\l)\Psi_\l)\otimes \Psi_\r,
\eeqa
where in the last step we made use of the fact that the sequence $ a\mapsto \int_{-a}^a ds\, \be^{(\l)}_{\t/\sqrt{2}}(B_{\l}^*B_{\l})\Psi_\l$
converges, as $a\to\infty$, in the norm topology of $\hilk_\l$ (cf. Theorem~\ref{H-A}) and  $t\mapsto (f_+^{\eta}(1-s/t)-1)$ converges to zero, as $t\to\infty$, uniformly in $s\in [-a,a]$.  
Equality~(\ref{detector-convergence-two}) is proven analogously.
As a consequence of (\ref{spectral-measure}), (\ref{detector-convergence-one}) and (\ref{detector-convergence-two}), we obtain
\beqa
& &Q^\tout_{\Ri}(B_\l\otimes I)Q^\tout_{\Le}(I\otimes B_\r) \PP(\De_{\eps}(e_\Ri,e_\Le))\hil_{\cc}\non\\
&=&Q(B_\l)\PP_\l([e_\Ri,e_\Ri+\eps])\hilk_{\l,\cc}\otimes Q(B_\r) \PP_\r([e_\Le,e_\Le+\eps])\hilk_{\r,\cc},
\eeqa
where we used Lemma A.2 of \cite{DT11}, as in the discussion after formula~(\ref{tensor-product-measures}) above.
To conclude the proof, it suffices to show that
\beqa
\PP_\l([e_\Ri,e_\Ri+\eps])\hilk_{\l,\cc}&=&\Span\{\,  Q(B_\l)\PP_\l([e_\Ri,e_\Ri+\eps])\hilk_{\l,\cc}\,|\, B_\l\in \LLL_{\l,e_+}\,\}^{\cl},  \label{one-dim-two}\\
\PP_\r([e_\Le,e_\Le+\eps])\hilk_{\r,\cc}&=&\Span\{\,  Q(B_\r)\PP_\r([e_\Le,e_\Le+\eps])\hilk_{\r,\cc}\,|\, B_\r\in \LLL_{\r,e_-} \,\}^{\cl}.
\eeqa
It is enough to prove the first equality above, as the second one is analogous. We proceed  similarly as in  the proof of Theorem~\ref{main-theorem-waves}:
By the translational invariance of $Q(B_\l)$ it is obvious that the subspace on the r.h.s.\! of (\ref{one-dim-two}) is
contained in the subspace on the l.h.s. Let us now assume that the inclusion is proper, i.e., we can
choose a non-zero vector $\Psi \in (\PP_1-\PP_0)\hil_{\cc}$, where $\PP_1:=\PP_\l([e_+,e_++\eps])$ and $\PP_0$ is
the orthogonal projection on the subspace on the r.h.s.\! of (\ref{one-dim-two}).
By Lemma~\ref{borchers}, there is an operator $B_\l\in\LLL_{\l,e_+}$ 
s.t.\! $B_\l\Psi \ne 0$. Then it is easy to see that $(\Psi| Q(B_\l)\Psi) \ne 0$
which means that $(\PP_1-\PP_0)Q(B_\l) \Psi\ne 0$. Hence  $Q(B_\l)\Psi \neq 0$ and $Q(B_\l)\Psi \notin \PP_0\hilk_{\l,\cc}$,
which contradicts the definition of $\PP_0$. \qed

\section{Completely rational conformal nets}\label{rational}
\setcounter{equation}{0}

In this section we consider particle aspects of completely rational conformal nets,
whose definition is summarized below. This class contains massless two-dimensional
theories in a vacuum representation which are not asymptotically complete in the
sense of waves. Nevertheless, as we show below, they have the property of generalized
asymptotic completeness. 

In the previous section we introduced the concept of a local net $(\A,\U)$ of von Neumann algebras 
on $\real$. Suppose that  $\A$ extends to a local net on the circle $S^1$ (understood as a one-point compactification
of $\real$) and $V$ extends to a unitary representation of the universal covering of the M\"obius group $\ov\Mob$, 
s.t.\! covariance still holds. Then we call the extension (resp. the original net)  a \bf M\"obius covariant \rm net on $S^1$ (resp. on $\real$). Similarly, 
if $\U$ extends to a projective unitary representation of the group of orientation preserving  diffeomorphisms of $S^1$, denoted
by $\diffs1$, 
s.t.\! covariance still holds and $\U(g)A\U(g)^*=A$ if $A\in \A(\I)$ and $g\in\diffs1$ acts identically on $\I$,
then we say that the extension (resp. the original net) is a \bf conformal \rm net  on $S^1$ (resp. on $\real$).

A conformal net   $(\A,\U)$  on $S^1$ is said  
to be {\bf completely rational} \cite{KLM01} if the following conditions hold:
\begin{enumerate}
\item {\bf Split property.} For intervals $\I_1,\I_2 \subset S^1$, where
$\overline{\I_1} \subset \I_2$, there is a type I factor $\mathcal F$
such that $\A(\I_1)\subset \mathcal{F} \subset \A(\I_2)$.
\item {\bf Strong additivity.} For an interval $\I$ and $\I_1,\I_2$ which are made from $\I$ by
removing an interior point, it holds that $\A(\I) = \A(\I_1)\vee \A(\I_2)$.
\item {\bf Finite $\mu$-index.} For disjoint intervals $\I_1,\I_2,\I_3,\I_4$ with a clockwise (or counterclockwise)
order and with the union dense  in $S^1$, the Jones index of the inclusion
$\A(\I_1)\vee\A(\I_3) \subset \left(\A(\I_2)\vee\A(\I_4)\right)'$ is finite.
\end{enumerate}
Among the consequences, we recall that a completely rational net $\A$ has only
finitely many sectors and any (locally normal) representation of $\A$
(on a separable Hilbert space) can be decomposed into a direct sum of irreducible representations \cite{KLM01}.

Now let $(\B,U)$ be a local net of von Neumann algebras on $\RR^2$ in a vacuum representation.
$(\B,U)$ is said to be {\bf M\"obius covariant} if the representation $U$ of  translations extends to the group
$\overline\Mob\times\overline\Mob$ and the covariance still holds in the sense of local action
(see \cite{BGL93}). 
If $U$ further extends to a projective unitary representation of
the group $\diffs1\times\diffs1$ which acts covariantly on the net, and it holds that
$U(g)AU(g)^* = A$ if $A \in \B(\O)$ and $g\in \diffs1\times\diffs1$ acts identically on $\mco$, then the net $\B$ is said to be \bf conformal\rm.
See also \cite{KL04-2} for a general discussion on conformal nets on two-dimensional spacetime.

We define subgroups $\widetilde{G}_\l := \overline\Mob\times \{\iota\} \subset \overline\Mob\times\overline\Mob$
and $\widetilde{G}_\r := \{\iota\}\times\overline\Mob \subset \overline\Mob\times\overline\Mob$,
where $\iota$ denotes the unit element in $\overline\Mob$.
Following \cite{Re00},  for any interval $\I \subset \RR$,
we introduce the von Neumann algebra $\A_\l(\I) = \B(\I\times \J) \cap U(\widetilde{G}_\r)'$. This definition
does not depend on the choice of $\J$, since the group $\overline\Mob$ acts transitively
on the set of intervals. Analogously, one defines $\A_\r(\J) := \B(\I\times\J)\cap U(\widetilde{G}_\l)'$.
In this way  we obtain two families of von Neumann algebras parametrized by intervals contained
in $\RR$. It was shown by
Rehren  that both $\A_\l$ and $\A_\r$ extend to M\"obius covariant nets on the circle $S^1$
\cite[Section 2]{Re00}. If the net $(\B,U)$ is conformal, then both chiral components $\A_\l$ and $\A_\r$ are nontrivial.
Indeed, they include the net generated by the diffeomorphisms of the form $g_\l\times \id$
and $\id\times g_\r$, respectively. Such nets, generated by diffeomorphisms, are called the
Virasoro (sub)nets.

We say that a conformal net $(\B,U)$ on $\RR^2$ is {\bf completely rational} if its chiral components
$\A_\l,\A_\r$ are completely rational. From the two nets $\A_\l$ and $\A_\r$ we can construct the
chiral net $\A_\l\otimes\A_\r$ as in the previous section, which can be naturally identified
with a subnet of $\B$. It is easy to see that the inclusion $\A_\l(\I)\otimes\A_\r(\J) \subset \B(\I\times \J)$ is irreducible
(namely, the relative commutant is trivial). Indeed, any element in $\B(\I\times\J)$ commutes with
diffeomorphisms supported outside $\I\times\J$, which are contained in $\A_\l(\I')\otimes\A_\r(\J')$,
where $\I'$ denotes the interior of the complement of $\I$ (in $\RR$ or in $S^1$, which does not matter thanks to the
strong additivity).
By the strong additivity, an element in the relative commutant must commute with any diffeomorphism. Hence it must be a multiple of the identity, since $\B$ is in a vacuum representation. From this and the complete rationality, 
it follows that the Jones index of the inclusion $\A_\l(\I)\otimes\A_\r(\J) \subset \B(\I\times \J)$ is finite
\cite[Proposition 2.3]{KL04-1}. Thus the natural representation $\pi_\B$ of $\A_\l\otimes\A_\r$ on
the Hilbert space $\H$ of $\B$ decomposes into a finite direct sum of  irreducible representations.

If $\A_\l$ and $\A_\r$ are both completely rational, then any irreducible representation of the
chiral net $\A_\l\otimes\A_\r$ is a product representation
\cite[Lemma 27]{KLM01}. From this it follows that if $\B$ is completely rational,
then the Hilbert space $\H$ can be decomposed into a direct sum
of finitely many product representation spaces of $\A_\l$ and $\A_\r$.
Thus the representation of the Virasoro subnets decomposes as well.
The representation $U$ of the spacetime translations can be obtained from
local diffeomorphisms, hence any representative $U(t,\xb)$ is contained in
$\bigcup_{\I\times \J} \A_\l(\I)\otimes\A_\r(\J)$. 
According to the decomposition $\pi_\B = \bigoplus_i \pi_i$ of the natural inclusion representation of
$\A_\l\otimes\A_\r$, $U$ is decomposed into a direct sum $\bigoplus_i U_i$
and each $U_i$ implements the translations in the representation $\pi_i$.
In other words, we obtain a decomposition of $U$ which is consistent with the above decomposition
of $\H$.

In the previous sections we saw that any product representation of a
chiral net is asymptotically complete in the sense of
Definition \ref{Asymptotic-completeness}.
It is easy to check that the direct sum of asymptotically complete
representations is again asymptotically complete. Thus we obtain:
\begin{theoreme}\label{th:ac-in-crnet}
Any completely rational net represented on a separable Hilbert space is asymptotically complete in the sense of 
Definition \ref{Asymptotic-completeness}.
\end{theoreme}
Recall that a two-dimensional conformal net is asymptotically complete in the sense of waves
if and only if it coincides with the chiral net  $\A_\l\otimes\A_\r$ \cite[Corollary 4.6]{Ta11}.
For a non-trivial
extension of a chiral net (see \cite{KL04-2} for examples and a classification result of
a certain class of conformal nets) asymptotic completeness in the
sense of waves fails, but generalized asymptotic completeness remains valid in the completely rational case
in view of the above theorem.

\appendix
\section{Auxiliary lemmas}

\bel\label{lm:translation-invariance} Suppose $Q^{\tout,\eta}(B_+)=\lim_{T\to\infty} Q^{T,\eta}_{\pm}(B_{+})$ exists on vectors from $\mathcal D$.
Then  $Q^{\tout,\eta}(B_+)$ is invariant under spacetime translations.
\eel
\proof Invariance under time translations is a consequence of time-averaging. We check invariance 
under space translations. Let $\Psi_1,\Psi_2\in \mathcal D$:
\beqa
& &|(\Psi_1|\big(Q^{T,\eta}_{+}(B_{+}) -  Q^{T,\eta}_{+}(B_{+})(\yb)\big) \Psi_2)|\non\\
& &\leq\int dt\,h_T(t) \int\, d\xb\, |f_{+}^{\eta}(\xb/t)-f_{+}^{\eta}((\xb-\yb)/t)| |(\Psi_1| (B^*_{+}B_{+})(t,\xb)\Psi_2)|\non\\
& &\leq C\int dt\,h_T(t) \, \sup_{\xb\in \real} |f_{+}^{\eta}(\xb/t)-f_{+}^{\eta}((\xb-\yb)/t)|,
\eeqa
where in the last step we made use of the fact that $Q(B_+)E(\De)$, defined in Proposition \ref{detectors-on-waves}, is a bounded operator for compact $\De$ by (\ref{harmonic-analysis}). Now we note
\beqa
\sup_{\xb\in \real} |f_{+}^{\eta}(\xb/t)-f_{+}^{\eta}((\xb-\yb)/t)|&=&\sup_{\xb\in \real} |f_{+}^{\eta}(\xb)-f_{+}^{\eta}(\xb-\yb/t)|\non\\
&\leq &\int_0^{\yb/t}d\yb' \, \sup_{\xb\in \real}|\partial f_{+}^{\eta}(\xb-\yb')| \leq C/t,
\eeqa
where in the last step we made use of the fact that $\partial f_{+}^{\eta}$ is non-zero only on a compact set. \qed

\bel\label{technical-commutators} Let $g_-^{\eta}\in L^{\infty}(\real)$ be  supported in $(-\infty, \eta]$ for some $0<\eta<1$ and $g_+^{\eta}(\xb):=g_-^{\eta}(-\xb)$. Let 
 $B\in\mathcal L_{\Ri,\de}\cup \mathcal L_{\Le,\de}$ and $F_{\pm}\in \mfa$ be s.t. $\Sp^{F_{\pm}}\al$ are compact. Let
\beqa
R^{T,\eta}_{\pm}(B):=\int dt\,h_T(t) \int\, d\xb\, g_{\pm}^{\eta}(\xb/t)  (B^*B)(t,\xb).
\eeqa 
Then, for any two compact sets $\De, \De'\subset \real^2$ 
\beqa
\lim_{T\to\infty}\|\PP(\De)[R^{T,\eta}_{\pm}(B), F_{\mp}(h_T)]\PP(\De')\|=0. 
\label{commutator-vanishing-one}
\eeqa
\eel
\proof We will show (\ref{commutator-vanishing-one}) only in the case involving the commutator $[R^{T,\eta}_{-}(B), F_{+}(h_T)]$, as the remaining case is analogous.
Let us first assume that $F_+$ is almost-local. In this case the argument is similar to the proof of Lemma A.3 of \cite{DT11}: By analogy to formula~(A.13) of \cite{DT11} we can write
\beqa
& &\|\PP(\De)[R^{T,\eta}_-(B), F_+(h_T)]\PP(\De')\|\non\\
& &\ph{444444444}\leq \!\!  \|g_-^{\eta}\|_{\infty} \int dt dt_1\, h_T(t) h_T(t_1)\!\! \int_{\xb\leq \eta t} d\xb\, \|[ (B^*B)(t,\xb), F_+(t_1,t_1)]\|.
\label{first-step-comm}
\eeqa
Let $L_T$ denote the l.h.s. of (\ref{first-step-comm}). Following the steps (A.13)-(A.16) of \cite{DT11} we  obtain that
\beqa
L_T\leq \|g_-^{\eta}\|_{\infty} \int dt dt_1\, h_T(t) h_T(t_1)\!\! \int_{\xb\leq \eta t} d\xb\, \chi(|\xb-t_1|\leq |t-t_1|+2r)+o(1), \label{L-T-integral}
\eeqa
where $\chi$ is the characteristic function of the corresponding set, $r=(1+\fr{1}{4}|\xb|)^{\eps}+T^{\eps}$, $0<\eps<1$ appeared in the definition of $h_T$  and $o(1)$ denotes a term which tends to zero as $T\to\infty$. The discussion below (A.16) of \cite{DT11} leads to the conclusion that the integrand is zero unless $|\xb-T|\leq c_3 T^{\eps}$, $c_3\geq 0$.  Now the restriction on the region of integration   gives
in addition $\xb\leq \eta (cT^\eps+T)$ for some $c\geq 0$. It is easy to see that these two conditions cannot be simultaneously satisfied for $0<\eta<1$ and arbitrarily large $T$, 
so we get $\lim_{T\to\infty}L_T=0$. This proves~(\ref{commutator-vanishing-one}) in
the case of almost-local operators $F_+$. 

In general, we choose a sequence of local operators $F_{+,n}$, s.t.\! $\lim_{n\to\infty}\|F_{+,n}-F_+\|=0$.  Since $\Sp^{F_{+}}\al$ is compact, we can choose a function
$f\in S(\real^2)$ s.t.\! $\supp\tf$ is compact and $F_+=F_+(f):=\int_{\real^2} dx\,F_+(x)f(x)$. Then $\lim_{n\to\infty}\|F_{+,n}(f)-F_+\|=0$
and, making use of relation~(\ref{harmonic-analysis}), we can replace $F_+$ with $F_{+,n}(f)$ in  (\ref{commutator-vanishing-one})
at a cost of the  following error term
\beqa
& &\|\PP(\De)[R^{T,\eta}_{-}(B), (F_+-F_{+,n}(f))(h_T)]\PP(\De')\|\non\\
& &\ph{44444444444}\leq 2\|g_-^{\eta}\|_{\infty}\|Q(B)\PP(\De'')\|\,\|F_+-F_{+,n}(f)\|,
\eeqa 
where $\De''\subset \real^2$ is a compact subset.  Here we made use of relation~(\ref{Arveson}) and of the fact that  $(F_+-F_{+,n}(f))$ has
compact  Arveson spectrum, uniformly in $n$. Clearly, this term tends to zero
as $n\to\infty$ uniformly in $T$. Since $F_{+,n}(f)$ are almost-local and  $\Sp^{F_{+,n}(f)}\al$
are contained in $\supp\tf$, (\ref{commutator-vanishing-one}) follows. \qed


\end{document}